\documentclass[acmsmall]{acmart}
\AtBeginDocument{%
  }

\begin{document}

\title{Generative AI and Creative Work: Narratives, Values, and Impacts}

\author{Baptiste Caramiaux}
\email{caramiaux@isir.upmc.fr}
\affiliation{%
  \institution{CNRS, Sorbonne Université, ISIR}
  \city{Paris}
  \country{France}
}

\author{Kate Crawford}
\affiliation{%
  \institution{Microsoft Research}
  \city{New York City}
  \country{USA}
}

\author{Q. Vera Liao}
\affiliation{%
  \institution{Microsoft Research}
  \city{Montreal}
  \country{Canada}
}

\author{Gonzalo Ramos}
\affiliation{%
 \institution{Microsoft Research}
 \city{Redmond}
 \country{USA}
}

\author{Jenny Williams}
\affiliation{%
  \institution{Microsoft}
  \country{USA}
}

\renewcommand{\shortauthors}{Caramiaux et al.}

\begin{abstract} 
  Generative AI has gained a significant foothold in the creative and artistic sectors. In this context, the concept of creative work is influenced by discourses originating from technological stakeholders and mainstream media. The framing of narratives surrounding creativity and artistic production not only reflects a particular vision of culture but also actively contributes to shaping it. In this article, we review online media outlets and analyze the dominant narratives around AI’s impact on creative work that they convey. We found that the discourse promotes creativity freed from its material realisation through human labor. The separation of the idea from its material conditions is achieved by automation, which is the driving force behind productive efficiency assessed as the reduction of time taken to produce. And the withdrawal of the skills typically required in the execution of the creative process is seen as a means for democratising creativity. This discourse tends to correspond to the dominant techno-positivist vision and to assert power over the creative economy and culture. 
\end{abstract}





\maketitle

\section{Introduction}\label{sec1}

The spread of generative artificial intelligence (generative AI) in the creative and artistic sectors is generating intense debate among academics, journalists, artists, policy makers and the general public. These conversations raise fundamental questions about art and human creativity, tackling subjects of cultural and philosophical interest, as well as questions about the future of the creative and artistic sectors~\cite{caramiaux_ai_2019, european_commission_opportunities_2022, world_economic_forum_impact_2018}. In parallel, the economic incentives surrounding AI in the creative industries further shape both public discourse and perceptions of artistic creation. This article examines the overlooked aspects of AI narratives in the arts, with a particular focus on how advances in generative AI technology and its implication in creative practices contribute to narratives about creative work. 

Narratives are cultural artifacts: stories that convey particular points of view or sets of values~\cite{bal_narratology_2009}. Narratives accompany, and sometimes drive, technological innovation as an element of communication for various audiences, including the general public, promoting cultural views and fostering its adoption~\cite{turner_counterculture_2010}. The stories create a bridge between the complex world of science and technology and the world of cultural representation of these technologies. These narratives become an important vector of adoption or rejection for people. For example, fictions around AI fuel existential fears that are taken up by actors in the field~\cite{cave_hopes_2019}, particularly in a number of companies and think tanks, and have a significant impact on innovation and policy making~\cite{gebru_tescreal_2024}.

When narratives touch on creative work and culture, they shape the way the general public perceives AI as well as the way they perceive artistic creation and creativity in general. And more importantly, the narratives used to convey its use in this environment, such as accessibility for all and effortless creativity, can have major consequences for culture in our contemporary societies. In this article, we investigate what constitutes the dominant narratives that are conveyed about generative AI technology in the fields of creativity and artistic creation. Through the analysis of these narratives, we investigate the following questions: Since narratives are told by different parties, \textit{who is telling the story of AI’s benefits for artistic creation? What explicit values do these narratives convey? What implicit assumptions underlie these values, and how do they influence or hinder creative work?}

In order to address these questions, we reviewed the narratives in online media publications that show up in relevant searches from popular search engines. The narratives are to be understood in the plural, as they arise from various intertwined perspectives. In these, however, we show that the voices are primarily not those of artists, but of journalists and tech actors. When artists have spoken on this topic, a different narrative emerges, one that is contrary to the dominant narrative from tech companies. We found that the discourse on creative work, when contextualised in AI, promotes the concept freed from its material realisation, such as craft skills, messy processes, markets. The separation of the idea from its material realisation is achieved by automation, which is the driving force behind productive efficiency. The idea takes precedence over the execution, which must be reduced to a minimum in terms of time. The withdrawal of the skills typically required in the execution of the creative process is then placed in the perspective of a democratization of creativity through the acquisition of rapid skills rather than long-term ones. This vision of creative work then creates tensions and a hierarchy in the economy of the creative sectors.  

This work contributes to research into the impact of generative AI on the creative and cultural sectors, in line with work alerting to the problems associated with injunction in the use of AI in art~\cite{jiang_ai_2023,donnarumma_against_2023}. We emphasize here that the dominant vision is a partial one, which is primarily not narrated by artists and which, in fact, reinforces artists’ fears and anxieties about this technology, which will eventually prevent them from using it in an inspiring and original way~\cite{jiang_ai_2023,zapata_who_2023}.

\section{Background}\label{sec2}

\subsection{AI Narratives}

AI is not a neutral technology or a neutral term. There are narratives accompanying AI, where narratives are understood as cultural artifacts of various kinds that tell stories and convey particular points of view or sets of values [3]. AI narratives have been carried by many means: fictional cultural artifacts such as science fiction writings and films, marketing slogans or product value propositions, tech articles in the media, interviews with product leaders, or cultural critiques.  

AI narratives have to be taken seriously because they can heavily influence public acceptance of the technology and they can also shape the way policymakers see the technology. The power of using the cultural dimension of AI was understood by tech companies, who, with a view to communicating with customers, rebranded machine learning (a discipline and set of techniques) as AI after the breakthrough in the field in the 2010s~\cite{whittaker_steep_2021}. The discourse aimed to communicate technological products as having major significance in terms of scientific innovation. Paradoxically, these predictive technology products are opaque and difficult to interpret, which has led to their behavior being likened to a form of magic. Campolo and Crawford~\cite{campolo_enchanted_2020} describe this phenomenon as “enchanted determinism,” whereby the perception of technology as magical and deterministic — thus outside of human control — allows companies to deflect accountability for the full impacts of their systems. In a recent article, Chubb et al. point out that dominant narratives are “polarized between notions of threat and myopic solutionism”~\cite{chubb_expert_2024}. In their article, the authors propose to explore the missing stories, especially about the everyday impacts of technology. These polarized narratives are fueled by fictions that often depict simplified stories and exemplify the fears and hopes of AI technology [11], which eventually mirror simplified emotions and values of our society. As Herman puts it: “[...] the desire to tell dramatic stories requires certain types of AI, for example humanoid robots or almighty systems. Thus, second, science-fictional AI is not necessarily about the technology but can be a metaphor for other social issues.”~\cite{hermann_artificial_2023} (p. 320). 

AI narratives also carry stereotypes. These stereotypes may, for example, be linked to how race and ethnicity are represented. In a recent study titled “Whiteness of AI”, Cave and Dihal~\cite{cave_whiteness_2020} argue that “AI is predominantly racialised as White because it is deemed to possess attributes that this frame [White racial frame] imputes to White people” (p. 696). They explain that there is a historical exclusion of non-men and non-whites from positions that are the best paid and considered professional, which is present in the way AI is represented~\cite{cave_whiteness_2020,noble_algorithms_2018}. Other recent works have shown how race should be considered, especially the role of people of color in the development and use of technology and the persistent raciale biases in contemporary technology~\cite{benjamin_race_2020}. The very choice of terms used has a territorialist and colonialist connotation: the name given to machine learning models has evolved from pre-trained model, foundation model to frontier model~\cite{gebru_changing_2024}. However, the paradox, as Alondra Nelson has shown, is that narratives tend to portray technology as erasing distinctions between race, gender or ability (the libertarian viewpoint), but end up promoting exclusive technology~\cite{nelson_introduction_2002}.

These studies show that technology is not neutral and that cultural studies of AI are important to shed light on its impacts. It is then surprising that few studies focus on this problem in the context of the creative and artistic sectors. These sectors are structuring the cultures of our societies. The narratives on technologies that have the capacity to transform these sectors then have the capacity to transform our cultures.

\subsection{AI in the Arts}

The use of AI in artistic practices has a long history. The Aaron software, created by artist Harold Cohen to produce drawings autonomously, is an example of the first experiments in integrating AI into artistic practice. Similarly, Vera Molnár notoriously used autonomous procedures to create repetitive patterns of deformable geometric shapes. In music, back in 1956, Lejaren Hiller first experimented with the first computers and implemented complex processes that have been automated to generate a musical score, called IIliac suite for String Quartet. Generative forms of music based on computer processes were then explored by numerous composers such as Daphne Oram with Oramics and Laurie Spiegel. In the 2000s, David Cope developed musical intelligence experiments. François Pachet, in France, developed musical algorithm based Markov chains as a composition and improvisation tool~\cite{pachet_continuator_2003}. 

The use of AI in artistic practice spread wider when first data-driven learning technologies were used to create visual art. In 2012, the deep learning breakthrough~\cite{krizhevsky_imagenet_2012} showed that Machine Learning (ML), a set of statistical methods able to perform prediction and generation based on learned structure from a dataset, has the capacity to generate images, sounds or texts that are creatively, and artistically interesting~\cite{akten_deep_2021}. While not specifically targeting artists and creatives, AI technology based on deep learning started to be appropriated by a small community of artists that experimented with newly released AI techniques~\cite{caramiaux_explorers_2022}. One of the first deep learning models used was the inception algorithm (or DeepDream), designed by Google engineers and artists~\cite{mordvintsev_inceptionism_2015}. AI Art then greatly proliferated after the publication of the Generative Adversarial Network (GAN) algorithm in 2014~\cite{goodfellow_generative_2014} and its extensions (see for instance, \cite{brock_large_2018,radford_unsupervised_2015}). This family of techniques made it possible to generate good-quality images, refine the model’s aesthetics by training it on a selection of curated images, and explore the abstract visual space learned by the model, creating surprises and interesting accidents for artists. These experiments started out as niches, often run by artists hacking code from Github. Then these experiments began to attract attention: an exhibition was organized at Google in 2015, and the Ars Electronica festival dedicated its annual theme to AI in 2017. In 2018, Christie’s auction house sold a digital artwork created using GAN. Communication started to be important: this artwork was publicized as the first work created by AI~\cite{daniele_ai_2019} and was sold several hundred times higher than the estimated price when the house first put the artwork up for sale. 

Media coverage of AI’s uses in the artistic and creative sectors expanded following more recent AI innovations. In 2020, innovations in the AI field demonstrated that larger ML models could mimic the structure of natural language, such as English, and generate plausible continuation of a provided sentence~\cite{brown_language_2020}. Similar model architectures were then used to learn relationships between text and image, allowing generating images based on inputs of textual descriptions~\cite{radford_learning_2021}. Shortly after, companies released text-to-image systems, such as OpenAI’s DALL-E\footnote{\url{https://openai.com/index/dall-e-2/ }}, MidJourney\footnote{\url{ https://www.midjourney.com }}, and Stable Diffusion\footnote{\url{https://stability.ai/news/introducing-stable-diffusion-3-5 }}. These systems were explicitly marketed  to creatives and artists. They offer image outputs with high levels of realism and polish, are easy to use and support versatile use cases, contributing to their popularity across diverse communities. By employing natural language as input, these systems provide a way to lower the technical barrier for content creation~\cite{chang_prompt_2023,sanchez_examining_2023}. Their simplicity of interaction and versatility mean they can be used for production as well as ideation~\cite{rajcic_towards_2024}. Prompting becomes part of this new artistic practice: prompts can be viewed as artworks in their own right, much like the model-generated outputs they produce, as well as tools by  artists who frequently employ prompt templates to craft these outputs~\cite{chang_prompt_2023}. In 2022, the cover of the general public magazine Cosmopolitan was produced using an AI-powered tool for generating images from text input. The magazine communicated that “The World’s Smartest Artificial Intelligence Just Made Its First Magazine Cover”\footnote{\url{https://www.cosmopolitan.com/lifestyle/a40314356/dall-e-2-artificial-intelligence-cover/ }}, , and “it only took 20 seconds to make”. This exemplifies the rapid integration of AI into mainstream media and underscores the growing role of AI in creative and commercial domains. And this was accompanied by a backlash from the artistic community (see, for instance, Steven Zapata’s presentation~\cite{zapata_end_2022}). Firstly, it became public that these new models were trained on databases containing copyrighted content, owned by the artists~\cite{xiang_artists_2022,baio_exploring_2022}. In addition, the assimilation of prompting practice with artistic practice was seen as a diminishing view of creativity~\cite{johnson_is_2023}.

AI artists are not only engaging with the technical aspects of AI, but also its cultural and political dimensions. Through their work, they are critiquing the culture of AI research, highlighting its focus on technical performance at the expense of social and ethical considerations~\cite{caramiaux_explorers_2022,kawakami_impact_2024}. They are also using their art to raise awareness about issues of bias, labor exploitation, and power dynamics inherent in AI systems~\cite{caramiaux_explorers_2022,hemment_ai_2023}. By making these issues visible, AI artists are contributing to a broader societal discussion about the responsible development and deployment of AI.

These latter works bring to the fore elements of discussion, on the part of artists, about what is involved in using AI in creative work. In this article, we question whether these elements can be found in the narratives around AI in art, when communicated to the general public through the media. In the following section, we present our methodology for investigating this question.

\section{Method}

In this section, we present the methodology used to answer our research questions on the study of AI narratives in the arts. In particular, we present the procedure used to build the corpus of articles and their analysis.

\subsection{Corpus creation}

We created a corpus of press articles, news, blog posts on AI in the arts. We took these materials from several online media outlets. Some of them specialize in technology or in art, while others address the general public. Collecting such a corpus was challenging, as there is no centralized database for finding articles dealing with AI in the arts or AI in the creative sector. Our approach has been to place ourselves in the role of people asking questions about the issue and making queries on their search engines to gather information. 

We started with two search engines that account for most queries in Europe and the USA, the parts of the world where this research was carried out. These search engines were Google and Microsoft Bing As request keywords we used: “[ Art AI $|$ Artist AI $|$ Creative AI ]” (where “$|$” means “or”) paired with “[ articles $|$ news $|$ posts $|$ innovations $|$ products $|$ companies ]”. This amounts to a total of 18 queries. The search was conducted systematically using both Bing and Google search APIs. Each query is specified to return responses whose date is between 2020 and Today. We chose to start in 2020 because we date the rise of generative AI, as it is understood today, to that year. This date also corresponds to the spread of generative AI in the creative and artistic sectors with the  release of several relevant products, for instance, enabling image generation from text description. 

As the list of search results to each query can be long (several hundreds of responses to each query), we chose to collect the first 90 search results  for each query, which means the first 9 pages in the standard interface of these two search engines. We assume that the articles beyond the first 9 pages receive significantly less visibility. Finally, we merged all the query responses and removed the duplicated results. As a result, we collected 1720 query responses. 

\subsection{Corpus filtering and analysis} 

\begin{figure}[!ht]
    \includegraphics[width=\linewidth]{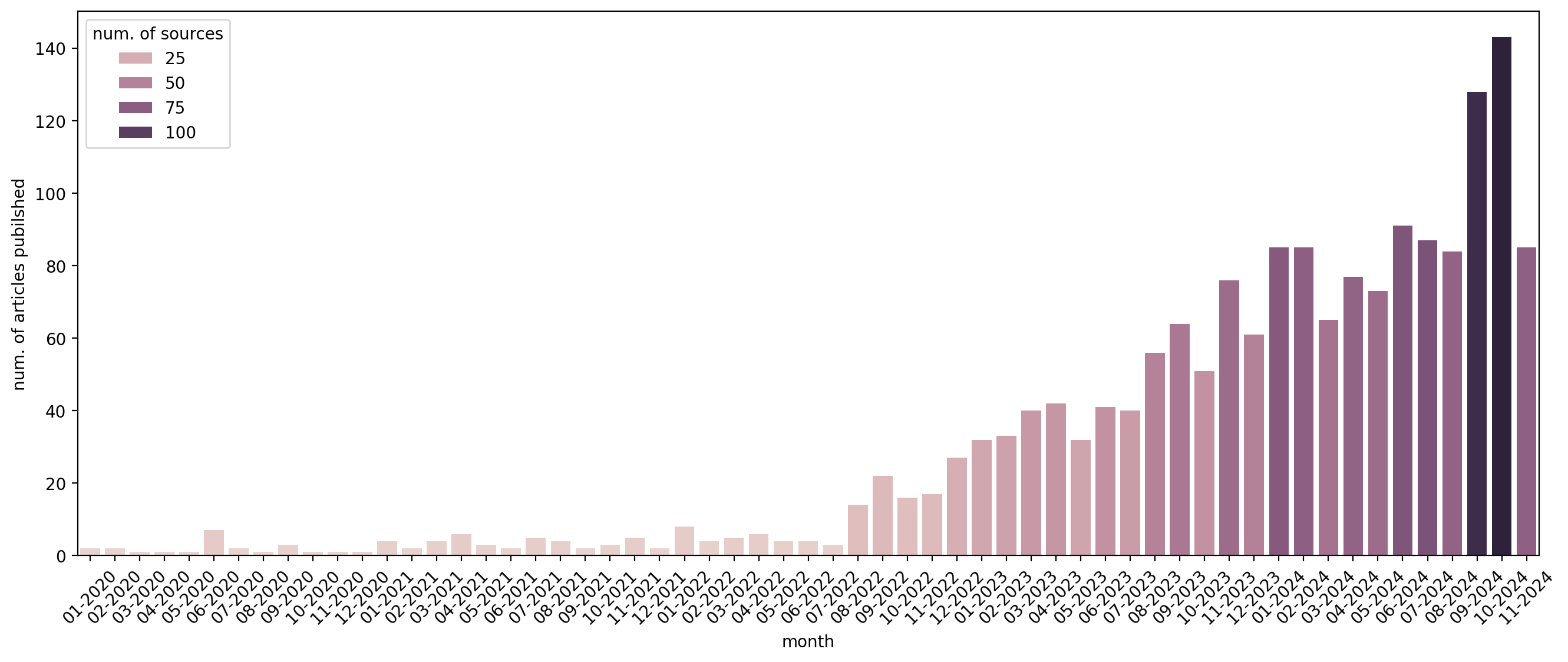}
    \caption{Bar chart of the number of publications per month obtained from our queries over the period January 2020 to the present date. The colours indicate the number of sources: a darker colour means a larger number of sources, while a lighter colour means a smaller number of sources.}
    \label{fig:barchart}
\end{figure}

The corpus includes web pages from many sources. Figure~\ref{fig:barchart} depicts the number of articles published and the number of sources per month over the last four years, showing an increase in the number of publications correlated with an increase in the number of sources. 

In this work, our aim is to identify a subset of representative online web pages that we will analyse qualitatively. We opt for this approach in order to provide a more granular analysis of our research question on the representation of creative work in online media and by creative technology companies. 
First, we filtered the data set in order to retain a subset of web pages from sources that had published several articles on the topics over the years. We removed sources that were only occasionally  published on the topic. We set the threshold at more than 5 articles over the 4 years. This threshold means more than one article per year (on average) and makes the list of final articles easier to handle for a qualitative analysis. In addition, we removed from the list academic publications and those related to tutorials on the use of AI tools in creative work. Finally, we included in the corpus websites from companies developing generative AI technologies for the creative and artistic fields. We extracted from these websites the general product descriptions. The final corpus comprises 188 articles from 19 sources plus 19 product descriptions. This corpus was analysed using inductive coding and thematic analysis. In this article, we present the part of the analysis relating to creative work. We produced a codebook containing 102 codes. These codes were then grouped into 5 themes that are presented in the following section. 

\section{Findings}

In this section, we present the results of our qualitative analysis. From the media outlets we identified five explicit values in the analysed narratives: Automation over manual work; efficiency over exploration; concept over execution; artifact over process; skills with a low barrier to entry over skills that take effort and time to learn. For each explicit value, we highlight the implicit narratives that underpin them and their impacts on creative work. Table 1 reports a summary of the findings, on which we elaborate below

\begin{figure}[!ht]
    \includegraphics[width=\linewidth]{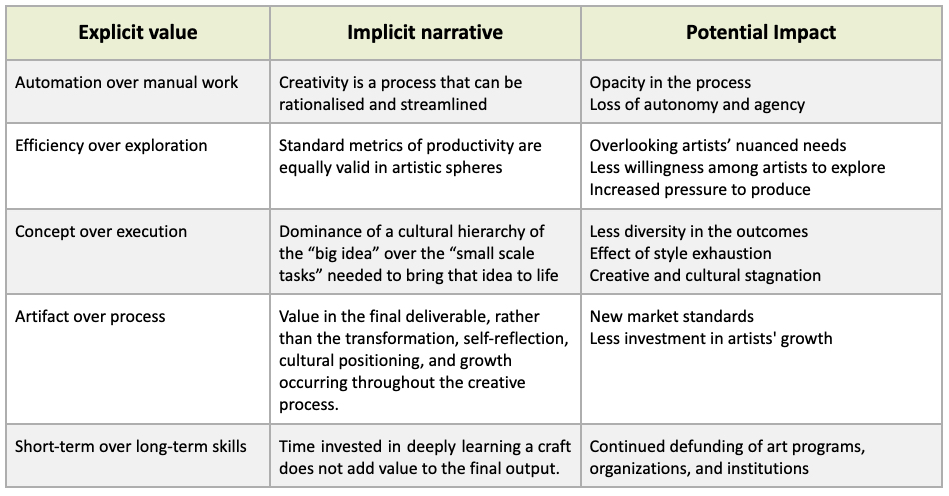}
    \caption{Summary of the findings. We identified five explicit values in the analysis (first column), from which we exhibit their implicit narratives (second column) and impacts (third column)}
    \label{tab:table}
\end{figure}

\subsection{Automation over manual work}

A 2023 article published in Wired started with the following sentence: “It used to be widely thought that creative work would be one of the last things to be automated. After 2022, some may reconsider.”~\cite{knight_where_2023} 2022 was the year where major text-to-image systems were released (Midjourney, Stable Diffusion, Dall-E, or Craiyon). These systems have been designed to automate the process that links the formalisation of a description in text to a visual rendering as an image: “The process of creating visuals is complex, manual and often requires specialized skills. [system] was created to address this challenge — providing a visual generative AI platform tailored to enterprises that digitizes and automates this entire process.”~\cite{wiggers_this_2023}

Certain artists such as Vera Molnar or Harold Cohen, mentioned above, used automation in their work. This was motivated by personal artistic gesture and aesthetics~\cite{molnar_toward_1975}. With generative AI, the scale is not individual but global. The motivation behind the automation of the creative process through the use of generative AI does not stem from individual creative or artistic needs. Rather, automation is rooted in the capitalist pursuit of efficiency, productivity, and profit maximization and has been a driving force of industry since the third Industrial Revolution~\cite{noble_forces_2017}. During the Industrial Revolution, mechanization replaced manual labor in factories, significantly reducing costs and increasing output~\cite{mowery_technology_1991}. This trend continued with the advent of electricity, assembly lines, and, more recently, digital technologies~\cite{xu_fourth_2018}. In today’s capitalist society, automation addresses the demand for faster production, and scalability in a globalized economy, while also reducing dependency on human labor. As put in a 2023 article published in FastCompany, “Generative AI tools are not artists; rather, they are creative factories. This is more like the industrial revolution for creativity. With these tools, creative efficiency is set to multiply astronomically.”~\cite{clugston_generative_2023}. In turn, the narratives promulgated by AI-driven creative projects often highlight the technological marvel of the artistic medium (e.g., machine learning algorithms) and obscure the labor that goes into making, hacking, and adapting the algorithms to make them work for bespoke artistic contexts~\cite{caramiaux_what_2022}. 

In addition, through the automation of the creative process, the underlying assumption is that creativity can be rationalised and streamlined such as becoming automated by an algorithm. Then, one of the dangers of this automation is the opacity of the underlying processes used to generate content, and the loss of artists' autonomy in relation to their object of work.

\subsection{Efficiency over exploration}

There is a dominant discourse that privileges the notion of productivity efficiency in creative processes. The notion of efficiency primarily highlights time as a critical factor to be reduced during the creative process in the discourse shaping AI culture in the creative and artistic domains. 

The depiction of creativity as a time-consuming, inefficient process is common in the narratives of companies developing AI-based technology solutions for artists and creatives. For instance, RunwayML, a company specialized in video editing, states that they “bring research into production within weeks instead of years.” Similarly, Captions, a company doing video storytelling, states that its “mission is to empower our customers, regardless of background or experience, with the ability to bring their vision to life in seconds.” Or AIVA, a European music generation company, proposes an assistant that “allows you to generate new songs in more than 250 different styles, in a matter of seconds.” The notion of efficiency implies that creating an artistic artifact or output involves a set of time-consuming tasks, and that those tasks can be  performed and controlled by the AI-powered system instead of by the artist. The Canadian AI video creation company Lumen5 exemplifies this value proposition: “Video editing is a time-consuming process because it requires a lot of small tweaks and fine tuning. Using Machine Learning, we automate these tasks to help our users create quality videos with minimal time, effort, and training.” 

This emphasis on efficiency cultivates the assumption that creative processes are not efficient enough (for the role assigned to them). In other words, standard metrics of productivity, such as quantity and speed, are equally valid in artistic spheres in order to assess the creative process. In doing so, technology providers state a problem that aligns conveniently with the technological solutions offered:  they promote their products as necessary solutions to enhance efficiency and productivity.  These companies may then be reinforcing an efficiency-focused framework that overlooks the nuanced needs of creative practitioners, shaping perceptions of the creative process to prioritize speed over exploration, mistakes and accidents. However, research has shown that artists value the latter even when engaging with AI: they often exploit the inherent unpredictability of AI algorithms, embracing chance and unexpected outcomes as part of the creative process~\cite{caramiaux_explorers_2022,hertzmann_can_2018}. This process is surely timely but it has also been shown to lead to the discovery of novel aesthetics, artistic styles, and forms of expression~\cite{chang_prompt_2023,jiang_ai_2023}. Consequently, by encouraging efficiency, these narratives bring the very likely prospect of discouraging artists from exploring, while increasing the pressure on production. As artist Mario Klingemann put it: “Rather than contributing more to the ever-growing pile of ‘content’, artists might need to create less. This seems counterintuitive, but in a world saturated with AI-generated output, the artist might shift towards curation, distillation, and extracting meaning from the noise.”~\cite{silva_mario_2024}.

\subsection{Concept over execution}

AI technologies aimed at artists, including text-to-image and text-to-video systems, place a strong emphasis on the role of conceptualization in the creative process, suggesting that a compelling idea holds greater value than its material realisation: the technical execution of that idea, or the embodied actions made to realize it. There is a dominant narrative that creativity can be let free through a streamlined path from the creator’s idea to the final artifact, be it an image, a sound, a video, a text, etc. For instance, Boomy (music generation), Leonardo.AI (image generation), Synthesia (video creation), all use the metaphor of creativity being constrained and that their products would help to “unleash”. “Let your creativity flow, and make AI do the rest” is the marketing slogan of Descript, which specializes in video editing. 

Within the framework of generative AI, the stance is that creation occurs as a direct transformation from concept to artifact (output, or object). This process involves a disembodied approach to creativity, in which the importance of the mind and abstract conceptualization is put forward, over other aspects of production. This emphasis suggests that the creative act is about ideation and can be limited or constrained by material engagement. “AI cultivates and caters to our passivity, seeming to offer the fruits of creativity and self-examination without the effort and self-doubt.”~\cite{horning_word_2022} The idea is that all an artist has to do to make a creation is prompting, i.e., writing a sequence of words that defines a command for the generation algorithm. 

Commentators have linked the importance of the concept and the idea in the creative process to the postmodern and surrealist currents in art history. “A.I.’s use for rendering images follows a postmodern tradition in art, in which the final product is defined less by skill and more by the theory it posits”~\cite{pilat_i_2022}. Surrealists were using methods based on algorithms, chance, meditation methods, etc. These methods aimed at freeing themselves from a normative rationalism~\cite{dafoe_surrealism_2024}. Paradoxically, outsourcing a significant proportion of creative decisions to AI can lead, on the contrary, to the artist being constrained by the (unseen) current and past cultural motifs represented in AI training data, without going beyond them. As Ted Chiang explains, “If an A.I. generates a ten-thousand-word story based on your prompt, it has to fill in for all of the choices that you are not making. There are various ways it can do this. One is to take advantage of the choices that other writers have made, as represented by text found on the Internet; that average is equivalent to the least interesting choices possible [...]. Another is to instruct the program to engage in style mimicry, emulating the choices made by a specific writer, which produces a highly derivative story. In neither case is it creating interesting art.”~\cite{chiang_why_2024}

Delegating “small-scale” tasks to AI might not be an arbitrary framing. Companies emphasizing efficiency assume that these types of tasks are uninteresting or simply “busy work.” Yet these tasks might be considered by creatives as crucial elements of the creative process. Chiang argues exactly this point when he states that “...art requires making choices at every scale; the countless small-scale choices made during implementation are just as important to the final product as the few large-scale choices made during the conception. It is a mistake to equate ‘large-scale’ with ‘important’ when it comes to the choices made when creating art; the interrelationship between the large scale and the small scale is where the artistry lies.”~\cite{chiang_why_2024} The inclination of considering small adjustments, especially those that involve repetitive actions, as  less significant within the broader creative process may stem from a cultural hierarchy that places higher esteem on the conceptual and unique aspects of art, as in postmodernism, while undervaluing the technical precision and skill associated with craftsmanship. As Anna Ridler, conceptual artist working with AI put it: “I actually think that there are a lot of parallels between craft vs art and dataset vs algorithmic output. I think there are a couple of things like the relation between craft which is anonymous and less well regarded, and how it’s repetitive, versus art.”~\cite{caramiaux_explorers_2022} In AI art, Ridler explicitly exhibits the craft of her work by exhibiting the dataset she created, made of thousands of images, alongside with the model’s outputs, generated from this dataset. 

One of the dangers we see in the emphasis of concept over execution is a reduction in the diversity of content created, since a significant number of elements and details will be left to the decision of the generative model. This phenomenon starts to be documented~\cite{doshi_generative_2024} and can lead to stylistic exhaustion: “culture is going nowhere fast. The worries that accompanied artificial intelligence in 2023 reaffirmed this fear: that we’ve lost something vital between our screens and our databases, that content has conquered form and novelty has had its day. If our culture has grown static, then might we call our dissembling chatbots and insta-kitsch image engines what they are: mirrors of our diminished expectations?”~\cite{farago_i_2023}. 

\subsection{Artefact over process}

In the narratives that we analyzed, efficiency and productivity are understood as added value, which implies a specific definition of “value” based on the quantity of production rather than the quality of production and artistic experience and growth. Saving time increases productivity, i.e., the quantity of “things” produced by creative people and artists. As it is stated in a recent article published in Forbes, “For businesses, this will mean quicker project completion, while freelancers will fit more work into their schedules, allowing them to take on more work or offer lower prices to their clients”~\cite{marr_how_2024}. 

While this logic of increased supply driving down prices aligns with principles of capitalist economies, where competition and increase of offerings typically lead to lower costs, this dynamic raises significant questions regarding its desirability within the creative and artistic sectors. Creative outputs are significantly subjective and require time for ideation and refinement, aspects that are at risk in a model that values volume and speed over intrinsic quality. The Creative Economy already relies heavily on freelance and gig-based work, work that is marked by inconsistent income, limited protections, and potentially limited benefits~\cite{taylor_creative_2020}. The artistic labor market is known to attract more aspirant artists than the available jobs~\cite{menger_artistic_1999} partly due to an overestimation of the chance of success. The rise of AI in the creative sector may create higher incentives for aspiring artists, who could perceive careers traditionally characterized by autonomy and freedom as more accessible than ever before. This shift aligns with the increase of flexible art-related careers, where specialization may give way to adaptability across diverse opportunities~\cite{lingo_looking_2013}. The application of generative AI to creative and artistic work could further accelerate this trend, fostering a professional landscape that values versatility over narrow expertise. 

One of the consequences with these new AI advancements is also the risk to see generative AI establish new lower standards for quality, time to deliver, and pricing. As AI tools continue to evolve, they offer rapid, nearly free or cost-effective alternatives for tasks traditionally handled by creatives and that usually take time (as presented previously). This shift may redefine client expectations, prioritizing faster turnaround times and lower costs, which could marginalize creatives whose practices do not integrate easily with AI technologies or who lack proficiency in using these tools. For these individuals, staying competitive becomes increasingly challenging, as they may struggle to meet new industry benchmarks. Creatives who rely on traditional, manual, or non-AI-compatible methods may then find it difficult to follow their careers in an environment increasingly dominated by AI-driven production. This technological divide risks deepening inequalities within the creative sector. As artist Rubem Robierb stated in a CNBC article: the use of AI is “not a matter of choice”~\cite{handley_part_2024}. 

\subsection{Short-term over long-term skills}

One reason behind the focus on ideation over execution is to reduce the need to acquire time-consuming technical skills, which in return brings more people to use the tool. “I [Picsart’s CEO] really see it more like a copilot to allow humans to be even more creative and more advanced in their expression. So the things which were taking years of learning and training and fine-tuning, now you can do it easier and faster. It’s opened an opportunity for more people to be creative.”~\cite{snelling_why_2023} The creative process was seen as something previously reserved for artists, for those who had learned the technical skills to create. “Most of these users are not [...] professional artists, and that’s the point: They do not have to be.”~\cite{kelly_picture_2022}

The idea that it is now possible to create visual or musical works of art without any specific skills in drawing, color, or composition is being promulgated by many companies. For instance, Boomy proposes to “create original songs in seconds, even if you've never made music before.” Similarly, Lumen5, specialized in video creation, states that its “goal is to enable anyone without training or experience to easily create amazing videos in minutes.” As a final example, Photoroom, specialized in photo editing, “provides photo editing software powerful enough to create outstanding images yet simple enough to be used without any training.” These examples illustrate the underlying idea that it’s no longer worth spending time learning the skills required for drawing, video, or musical composition. The value is in the product.

Outsourcing important elements of the creative process begs the question of what skills artists retain or what skills they discard. AI is depicted as contributing to the “escalation of creativity” where “anyone can write at the level of Shakespeare, compose music with Bach, [and] paint in the style of Van Gogh”~\cite{pieters_creative_2016}.  World-renowned institutions such as the World Economic Forum have also published reports on the impact of AI in the Creative sector~\cite{world_economic_forum_impact_2018}, detailing how AI will slowly manage to perform increasingly complex creative tasks reserved for humans until now. The driving assumption in these projects is that algorithms can encapsulate creative knowledge and that by having access to suitable algorithms, artistic expertise can and will naturally emerge. 

These emerging narratives about the benefits of outsourcing expertise to generative AI echo the ones surrounding the global outsourcing of labor in the 1990s. Both sets of stories are grounded on the pursuit of efficiency, cost reduction, and scalability, often at the expense of traditional job security. In the 1990s, companies relocated manufacturing and service jobs to countries with lower labor costs, triggering debates about job displacement, economic inequality, and the erosion of local expertise. Similarly, today’s adoption of AI for tasks that previously required hard-earned skills do raise concerns about skill devaluation, and the ethical implications of automation.

Another vision is to take creative work to a macroscopic, generalist level, where artists would not have specific technical skills but would be artistic directors injecting ideas and their cultures into the automated process. “The next generation of artists and creatives will be closer to generalists. They’ll understand various aspects of art, anthropology, history, sociology, and music and use these broad cultural insights to guide the unrivaled dot-connecting-capacity of AI.”~\cite{clugston_generative_2023}. This last statement raises the question of future artist training. If technical skills lose their value, the training that enables them to be acquired is likely to decline.

\section{Discussion}

In this article, we have analyzed the dominant narratives around AI in the creative domain, focusing on the values that these narratives convey about creative work. We found that these narratives are anchored in five explicit values, which are: automation over manual work, efficiency over exploration, concept over execution, artefact over process, short-term over long-term skills. And each value embeds implicit narratives about what is creative work under the eye of generative AI and those who develop the technology. We believe that the impact of these narratives is not without consequences for the creative and artistic sectors, and that it is therefore important to analyze them in order to respond to them. 

First, the narratives surrounding generative AI in creative fields don’t necessarily present it as generating a decrease in creative work, but rather highlight its potential to increase and expand creative possibilities. The emphasis on automation and efficiency is aimed at empowering creatives by enabling them to focus on tasks considered rewarding. We believe that these narratives highlight generative AI as a tool for empowerment because they are rooted in a tradition where creativity is seen as problem-solving, suggesting that creative work follows rational and systematic processes that can be handled by generative AI~\cite{hsueh_what_2024}. This standpoint finds its root in the foundations of AI, as formalized by Simon: “I will try to show that there is no real boundary between WSPs [well-structured problems] and ISPs [ill-structured problems], and no reason to think that new and hitherto unknown types of problem solving processes are needed to enable artificial intelligence systems to solve problems that are ill structured.”~\cite{simon_structure_1973}. If creativity is seen as problem-solving, with its underlying processes that can be automated, generative AI becomes a valid tool and medium to assist creative practitioners and artists. However, the harms to creative work highlighted in this article, together with other criticisms made in recent research (e.g.~\cite{jiang_ai_2023}), show that creative work cannot simply be seen as problem solving and this view more certainly engenders problem-making solutions. 

Second, the notions of empowerment and democratization, highlighted in the narratives analyzed, have a long tradition as rhetorical means in the sphere of digital technologies (see for instance Turner’s work on cyberculture~\cite{turner_counterculture_2010}). However, tools that are said to be designed to promote inclusion and access may in fact reinforce existing inequalities or serve specific power structures~\cite{wijesiriwardena_is_2017}. In the early days of the Internet, researchers like Lisa Nakamura demonstrated the performativity of identities and stereotypes in spheres of online discussion, promoting radically democratic forms of discourse in which bodies and race do not intervene~\cite{nakamura_race_1995}. About digital divide, Alondra Nelson put: “Blackness gets constructed as always oppositional to technologically driven chronicles of progress.”~\cite{nelson_introduction_2002} In the creative field, tools are positioned as democratizing by the removal of specific skills and expertise that used to be necessary for creation, such as drawing, composition, visual literacy, and so on. We have shown that this framework can put additional tension on artists’ already precarious careers, and lead to a decline in support for artistic training. This use of “democratization” could deny creative and artistic careers to people who are not in a privileged position, and thus reinforce the exclusivity of access to these careers for an even more restricted part of those concerned. 

Third, the discourse around generative AI in the media outlets that we analyzed come from a majority of non-artists. Artists and creative practitioners are generally absent from the discourse, and their expertise is not always clearly represented in the communication surrounding the objects of their expertise. There are exceptions, such as Ted Chiang's article in The New Yorker~\cite{chiang_why_2024}. However, the vast majority of artists do not have access to these media. In this context, which marginalizes artists’ perspectives, we see another problematic effect, which is the invisibilization of the community. When the community is eclipsed, the focus shifts to individual users as the primary targets of these tools. Such an individualist view sees creatives as having desires regarding their tools, which can be taken in isolation of social influence (echoing Dewey’s description of individualism~\cite{dewey_art_1934}). Paradoxically, the design paradigms involved in generative AI, such as using natural language as inputs to creation, do not stem from artists’ practice and desires~\cite{epstein_art_2023,johnson_is_2023}. These design choices lower the entry barriers for a greater number of users who are neither experts nor artists, but amateur and apsiring artists. 

Finally, our work has its limitations. As technology evolves rapidly, we might wonder whether narratives don’t evolve too. For example, we believe that five years ago, narratives of generative AI in art emphasized the replacement of artists by technology, whereas current narratives focus more on augmentation and collaboration. Then, in this work, we have selected a subset of articles in order to conduct a qualitative analysis that is more granular and driven by our research questions than computational approaches. The process is therefore not systematic and calls on our subjectivity. This is both a strength and a limitation of the method.

% ... 

\bibliographystyle{ACM-Reference-Format}
\bibliography{bibliography}

\end{document}